# Working Paper

# Cost-Benefit Analysis of Kaptai Dam in Rangamati District, Chittagong, Bangladesh

Mohammad Nur Nobi*

**Abstract:**

This study aims to assess the net benefit of the kaptai dam on the Karnafuli river in Kaptai, Chittagong, Bangladesh. Kaptai Dam, the only hydroelectricity power source in Bangladesh, provides only 5% electricity demand of Bangladesh. The Dam is located on the Karnafuli River at Kaptai in Rangamati District, 65 km upstream from Chittagong. It is an earth-fill or embankment dam with a reservoir with a water storage capacity of 11,000 skm. Though the Dam's primary purpose is to generate electricity, it became a reservoir of water used for fishing and tourism. To find the net benefit value and estimate the environmental costs and benefits, we considered the environmental net benefit from 1962 to 1997. We identify the costs of Kaptai Dam, including its establishment cost, operational costs, the costs of lives that have been lost due to conflicts, and environmental costs, including loss of biodiversity, loss of land uses, and loss of human displacements. Also, we assess the benefits of electricity production, earnings from fisheries production, and gain from tourism to Kaptai Lake. The findings show that the Dam contributes tremendous value to Bangladesh. As a source of hydroelectricity, the Kaptai Dam is a source of clean energy, and its value might have been worthy of this Dam produced a significant portion of the electricity. However, providing less than 5% of the national demand for electricity followed by various external and sensitive costs, the Dam hardly contributes to the Bangladesh economy. This study thus recommends that Bangladesh should look for other sources of clean energy that have no chances of eco-political conflicts.

*Department of Economics, University of Chittagong, Bangladesh, Email: nurnobi@cu.ac.bd

**Acknowledgement**

The author is grateful to Planning and Development, University of Chittagong, Chittagong, Bangladesh for providing financial support to conduct this research.

# 1 Introduction

Kaptai Dam is the only hydropower source of Bangladesh located on the Karnafuli River at Kaptai in Rangamati District, which is 65 km upstream from Chittagong. It is an earth-fill or embankment dam with a reservoir (known as Kaptai Lake) with a water storage capacity of 11,000 skm. The primary purpose of the construction of the dam and reservoir was to generate hydroelectricity power. Started in 1957, the dam was built in 1962 with the assistance of USAID to the government of Pakistan. The total capacity of producing electricity of this dam is 230MW/Peak hour which is 5% of the national electricity demand. It has five units for electricity production, and the total cost of Unit 1, Unit 2, and a part of Unit 3 was Rs. 503 million, and the total cost of extension was Tk. 1,900 million. The government financed the ICA of the USA and DLF loan (Parveen and Faisal, 2002; Banglapedia).

The dam is 670 meters long and 45.7 meters wide. It has 16 gates for spillway on the left side of the main dam. The catchment area of the reservoir is 11,000 skm. The benefits of the dam were hydropower, significant flood control, irrigation and drainage, navigation, and enhanced forest resource harvesting. However, except for the irrigation and drainage, most of the benefits were served by the dam (Parvin and Faisal, 2002). In addition, it became the source of fish cultivation for a commercial reason and an attractive area to the visitors. As a result, it created many employments in the reservoir area.

The dam's construction submerged 655 skm areas, which included 220 skm areas (54,000 acres of cultivable land, which is equivalent to 40 percent of the cultivable land in the area) and displaced 18,000 families and 100,000 tribal people, of which 70% were Chakma. The dam also flooded the original Rangamati town and other structures. On the other hand, the local inhabitants who lost their homes and cultivable lands due to flooding were not compensated enough though there was a budget for the purpose. As a result, more than 40,000 tribals (mostly Chakma) migrated to India. Later on, those displaced and migrated people received military training with the help of the Indian government and organized a militant group known as Shanti Bahini (Kabir, 2005). Then, they started fighting against the country, and the conflict continued in the areas until there was a peace accord between the government of Bangladesh and Shanti Bahini in 1997. However, though the peace accord took place, one part of shanty Bahini continued fighting by changing their name to UPDF (United People's Democratic Front). Therefore, the country has to bear the extra cost of security in the hilly areas for a long time.

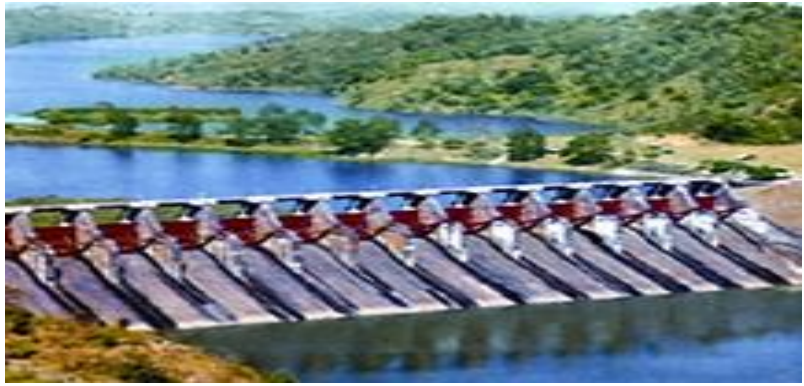

**Figure 1: Kaptai Hydro-electric Dam**

Source: Banglapedia, Date accessed February 28, 2018

(http://en.banglapedia.org/index.php?title=File:KaptaiHydroelectricityProject.jpg)

The dam construction also destroyed wilderness, environmental degradation, loss of wildlife and wildlife habitats (Banglapedia). So, after building the dam, that area became unstable in terms of politics, economy, and environment. Considering the issues mentioned above, this study aims to assess the net benefit of the kaptai dam; whether the dam on the Karnafuli river in Kaptai provides a net benefit to the country or not? To find the value of the net benefit, we considered an environmental net benefit. Therefore, the environmental net benefit is considered, whereas the project incurs environmental degradation and regular production costs. Therefore, in environmental cost-benefit analysis environmental cost of the project is considered in addition to the regular cost and benefit of the project. For estimating the environmental costs and benefits, we considered the time frame from 1962-1997. The dam production started in 1962 and then after conflict remained until 1997 when a peace accord was signed between the government of Bangladesh and the insurgent group; Shanti Bahini.

Kaptai Dam, the only hydroelectricity power source in Bangladesh, provides only 5% electricity demand of the country through all of its capacity cannot go for the operation now a day. Though the dam's primary purpose is generating electricity, it became a reservoir of water used for fishing and tourism. In contrast to its lifetime, the dam costs a lot, including the loss of cultivable land and housing of local people, environmental degradation, the value of lives lost (both in ethnic and plain land people), and extra security cost in the region to control the insurgency. Thus, considering all its costs and benefits, it is required to value the project whether it provides any net benefit to the national economy or not.

## 2. Objectives of the Study

The broad objective of this study is to find the Net Benefit of the Kaptai Dam considering the environmental issues. However, the specific objectives are to:

- Identify the costs of Kaptai Dam, including its establishment cost, operational costs, the costs of lives that have been lost due to conflicts, and environmental costs, including loss of biodiversity, loss of land uses, and loss of human displacements.
- Assess the benefits of electricity production, earnings from fisheries production, and gain from tourism to the Kaptai Lake,
- Find out the net benefits of this hydroelectric dam.

## 3. Literature Review

Kaptai hydroelectric plant is the primary source of electricity in Bangladesh. It is the lone hydroelectricity plant that supports renewable electricity in the country. The dependency on the dam for electricity supply was huge as only 3% of the people in Bangladesh had access to electricity after the independence in 1971(Islam and Khan, 2017). Though it was assumed that the electric plant produces 5% of the country's total supply of electricity, various research findings estimate various production in terms of changing demand of the country. For example, Akand et al. (2015) reported that the dam initially started with two hydropower units and produced 80 MW. However, currently, with five of its units, the dam produces 230 MW of electricity, meeting approximately 2.92% of the country's total consumption. After installation of its fifth generator, the total production reached the highest level at 230 MW in peak season in which 10-12 KW were used to meet local demand and the rest of the electricity were added to the national grid though the production decreases to 35 MW during the dry season (April-June) which can meet the demand for Kaptai Upazilla Sadar (Daily Star, 2007). According to the annual report 2017-18 of the Bangladesh Power Development Board (BPDP), only 1% of the total production is supported by the Kaptai Hydro plant (BPDB, 2017-18). Therefore, it could be assumed that the contribution of the Kaptai hydroelectric plant is no more significant in terms of its production.

Tourism is the indirect contribution of Kaptai Lake. For valuing the tourism contribution of this lake is estimated using the travel cost method for determining the influence of travel cost on-site visitation in Rangamati. So, the literature review section of the paper entails an in-depth discussion of other studies conducted applying the travel cost method. The reviews from these studies would help clearly understand the concepts and rightly analyze the results obtained in this research based on this methodology.

Richard et al. (1986) use the utility maximization recreation demand model to estimate the welfare obtained through recreational resources at a site and is linked with the travel cost method. For estimating the demand equation, a maximum likelihood (ML) procedure is undertaken. This helps in preventing biased estimates which might result from ordinary regression. The travel cost recreation demand equations have been derived from linear income as well as time utility functions. The aim is to find out the demand for recreation days for Lake Michigan's anglers and estimate the welfare obtained from fishing in Lake Michigan. The dependent variable in this study is the number of days the anglers spend fishing, while the primary independent variable is travel cost (which includes the opportunity cost consisting of monetary and time components simultaneously). Besides, the income variable and leisure time are used as two independent variables; indexes of angler's attitudes towards fishing and the ownership of boat as dummy have also been used. In the end, one best estimate for fishing in the Great Lake was obtained through OLS and ML estimators. The OLS regressions provide very biased outcomes, which ML did not. It was recommended that for the travel cost demand analysis, the dependent variable should better have data limits. The indexes were rightfully included in the study as they provide the correctly estimated equation significantly.

Lisa and Barry (1994) discuss how the travel cost method estimates the demand for hunting trips in Kansas. The data for this study is based on survey responses from 312 licensed Kansas hunters. An econometric approach is used for carrying out this travel-cost analysis, and the independent variables included here are the hunter's age, investment in hunting equipment, site quality characteristics, time-on-site for non-hunting activities, and length of stay. The educational qualification of the hunters, beauty, and success of hunting at the site was also included as dummy variables in this study. Here, the dependent variable is the visit rate defined here as the number of trips taken by the visitor in the last season, while the travel cost includes the cost of gas, food, and accommodation of the respondent's latest trip. It was obtained that if the travel cost increases, the visitation rate of the hunter decrease, which is shown by the negative coefficient of the estimated regression. Again, if the hunter invested more in the hunting equipment, the visitation rate will be higher. The hunter's age also influenced the visitation rate. In terms of the dummy variables, the education variable and the success of the hunting variable did not significantly influence the visitation rate. But if the hunter's assessment of the site was attractive, then it had a significant positive impact on the visitation rate. Using the travel-cost model and the above-mentioned variables, the visitation rate of hunters in many sites of Kansas was conducted. Finally, it was concluded that the hunters of Kansas had high demand of visiting at the recreational hunting sites, derived substantial welfare from it and so might be willing to bear the user access fees if needed. The implication here is that the policymakers should consider this fact while designing policies in the future.

John and John (1983) show a way of evaluation of a proposed site based on the travel cost method. They mentioned that "If a new site substitutes for an old one, then the willingness to pay for the new site substitutes (perfectly) for an old one, then the willingness to pay for the new site is the change in consumer surplus it creates under the demand curve for the old site. This should be distinguished from the total consumer surplus under a new demand function" (Hof 76). In case of multiple proposed sites, the system is more complex. The aim of this paper is to showcase an operational research framework in order to maximize/optimize the benefits and evaluate proposed site/sites based on this. A recreation allocation model (RAM) is developed where the choice variables are the amounts of recreation (in recreation visitor days) for a proposed site for individuals from different origins. The travel cost demand curves for the three wilderness areas of Colorado National Wilderness Preservation System were determined based on the data collected by the U.S Forest Service. An aggregate zonal travel cost was applied and the demand equations were constructed using a semi-log functional form. Finally it was concluded that optimizing the benefits derived from different proposed sites based on their substitutability with an existing site is simply an extension of the conventional travel cost model. Downward sloping demand functions were obtained and finally the conditions under which this model would be feasible have been discussed.

R. Craig Layman et. al. (1996) discuss how the travel cost method can be employed to conduct an economic valuation of the Chinook Salmon Sport Fishery of the Gulkana River under present and hypothetical fishery management conditions. The hypothetical management conditions are a doubled 1992 sport fish harvest, a doubled daily bag limit and a seasonal bag limit of five. Firstly, the travel cost method is used to estimate the economic benefits of the Chinook Salmon Sport Fishery of the Gulkana River with the existing management policies. Then we estimate the change in benefits due to the plethora of Chinook and changes in the sport fishing policies. Rather than using the contingent valuation method (CVM) like other economic studies, the hypothetical travel cost method (HTCM) was used to gain information about the hypothetical changes as well as to calculate the consumer's surplus in both existing and hypothetical situations. The data was collected from annual sport fish surveys from 1991 and 1992. In the econometric model, the number of visits is the exponential function of the travel cost and some other independent variables. The hypothetical conditions are used as the dummy variables in this case. The change in consumer surplus has been showed through two downward sloping demand curves with travel costs plotted on the x-axis and number of visits plotted on the y-axis. There was remarkable difference in the consumer surplus as the demand curve shifted rightward. After carrying out the OLS regression, it was found that after imposing the hypothetical conditions, the economic returns increased. It was also found that the anglers' preferences and tastes are at par with both actual and hypothetical trips simultaneously. Finally, some suggestions for conducting future studies have been recommended where the hypothetical scenario would be accepted in a test-retest structure.

Finally, Himayatullah and Rehana (2004) discuss about the study of economic valuation of Ayubia National Park (ANP) of the Himalayan Range Mountains by estimating the consumer surplus, visitors' willingness to pay and to find out whether an improvement in the recreational benefits would lead to a higher demand for park visitation. Here, both the travel cost method (TCM) and the contingent valuation method (CVM) have been used to assess the willingness to pay. Here, the variables include travel cost, income, substitute cost, education, age, household size, number of visits etc. It was found that people with higher incomes tend to visit parks more often. It was found that the yearly recreational value of the ANP is about 200 million. The consumer surplus generated here was Rs. 35.01 million.

From the above reviews of travel cost methods, it is easy to analyze the relevance of this approach in the context of economic valuation of Rangamati Lake. The research conducted on the tourism of Kaptai Lake using travel cost method could be a pioneer work since there is no significant study on the said issue in Bangladesh. Through this study, it is expected that the research gap in this particular field of study in this specific area will be eliminated.

The cost benefit analysis on the feasibility of the project is being currently assessed on Batang Ai Hydroelectric Plant (HEP). Wong et al. (n.d) have introduced Sarawak Energy Berhad's approach towards the mid-life renovation and renewal plan for this project. They have identified the problems related to unit subsystem and developed a working plan for the refurbishment; implementing active water and dispatch management and upgrading the turbine and generator for Batang Ai HEP to improve generator conversation efficiency. Most of the estimated costs include the purchase of raw materials for building and engagement of technical advisors.

As a reservoir, the Kaptai Lake serves breeding place of the fish species. The A study (2007) reported that 74 freshwater fish and 2 prawn species are available in the lake. After five years of formation of the lake the commercial fishing started in Kaptai Lake in 1965. At the beginning the carp variety dominated the Kaptai Lake amounting 81% of catch but currently, the carp variety decreased to 5% only, whereas small varieties are dominating catches by 95% (the Business Standard, 2020). All species bread naturally. In the fiscal year 2007-2008 there were 8250 MT of fishes were produced in the lake water. Ahmed et.al., (2006) cited that 73 fish species under 47 genera, 25 families including 2 species of prawn and one species of dolphin were detected in Kaptai reservoir. They also depicted that the annual average production of the fisheries is 3530 mt with an increase of 3.5% annually.

After installation of the dam the hills became instable politically and one group of the ethnic community started fighting against the government of the Bangladesh as militant. As a result, security issue became a concern for the government which borne additional costs.

Parveen and Faisal (2002) discusses the impacts of the Kaptai dam, in the Chittagong Hill Tracts of Bangladesh, on the tribal communities of that area in their study. The Kaptai dam was supposed to provide benefits in terms of hydropower, flood control, irrigation and drainage, navigation and enhanced forest resource harvesting. Most of these objectives have been served in various degrees except irrigation and drainage. More recently, commercial fish culture and recreation activities have been introduced in the lake. There was virtually no consultation with the tribal people in the 1960s during construction of the Kaptai dam. Now the consequences can be seen: both the Bengalis and the tribal people have paid a high price for the problems created by the dam and the displacement of 100 000 people. The outcome of the dam developments was intensive agriculture both in the remaining plain lands and in the hills, leading to soil erosion, productivity loss and water pollution caused by increased use of fertilizer and pesticides. Recently, the Power Development Board (PDB) of Bangladesh has announced a plan to install two new 50 MW units that will bring the capacity of the dam to 330 MW. This plan will cause the reservoir water level to rise and may take away about 7500 ha of the fringe land, which the tribal people use for rice cultivation during the April–August period in each year. Security became a concern in Chittagong Hill Tracts area as the conflicts and insurgency arouse because of the violation of the land rights of the tribal community (). As a result, the Government of Bangladesh deployed military forces to maintaining law and order in the region and about 30,000 troops were deployed there (The Report of the Chittagong Hill Tracts Commission 1991; cited from Nayak, 2019). Therefore, tribals were under control with the help of army but they were terrorized (Nayak, 2019). This deployment of army in the hilly region increased the cost of the government.

Different studies have been taken place in regards to the economic cost of hydroelectric dam in different regions. Some relevant literatures are reviewed here according relevancy to this research.

A study conducted by Oruonye (2015) examined the socioeconomic impacts of Kashimbilla multipurpose dam. The primary data were generated from field observation and interviews, while the secondary data was generated through secondary desk review of existing relevant materials. Some of the socioeconomic impacts of the dam include displacement of several communities and creation of two resettlement camps. The study findings show that despite the fact that the local communities were not involved in the dam project construction and resettlement. More than 200 communities stand to be affected by the Kashimbilla multipurpose dam construction. Observations show that the construction of the dam has been responsible for the relocation of large numbers of people from Birama. Most of the people moved to villages such as Hanki, Mango, Alahu and Tandun among others. Other impacts include loss of farmlands and historical areas used for Takaciyawa festival. It was reported that the dam led to the emergence of some diseases among the people of Jinagbanshin, Lukpo, Shibon Igba and Bariki Lisa as a result of the impoundment of the river. Some positive results include employment opportunity to 1,500 people in the area which is expected to increase to 2,000 after completion. Many construction works are springing up in the neighborhood of the dam. The Kashimbilla dam is also design to accommodate tourism. Special sections have been designed for this purpose that has not been completed at the moment. It is predicted that the dam might lead reduction of species diversity; loss of rare flora and fauna species. Impoundments lead to decreased woodland thereby adversely affecting wildlife communities. This leads to decreased hunting and associated uses. Medicinal herbs will also be lost. Furthermore, decreased forest plant communities lead to decreased timber production and attractiveness of an area to recreationists.

A study of Ngyuen et al (2017) in Vietnam stated that though the hydroelectric dam is benefiting, the economic cost of it especially the cost of displacement is overriding the benefits. Before the dam, the total land area was 168 ha which was reduced to 94 ha after the dam. People, who got displaced due to building of dam, lost their lands to produce rice dry crops and so on. This reduction resulted in less supply of dry crops in the market. Even supply of fish was reduced to 653kg from 6430kg due to displacement of people from the riverside area. Cattle production being not a major livestock is only seen in 15% households whereas it was 21% before the resettlement. There was also decline in raising chickens and ducks which resulted in lower market supply of chicken and ducks and failed to fulfill the market demand. This situation resulted in higher price of commodities and made people worse off. Being away from the riverside location, harvesting Lç Ô bamboo, honey, and rattan by travelling up river into protected forest have been disturbed and decreased. The loss of productive bamboo plantations and movement away from forest land has had a negative impact on income structure for households. On one hand people were losing on their livelihood and on the other hand the commodity price was on peak. Thus this situation made the economic condition worse off than it was before resettlement.

Similar case was found by Ahsan and Ahmad (2017) in their study of Bakun Dam in Malaysia. Almost 10,000 indigenous people were displaced for this project which resulted in several hardships. Most of them failed to secure a job in their new location. They were no longer able to continue their agricultural or hunting activities prior to the displacement. The interviewees of this research have stated that though government gives them water supply and electricity facilities which they didn't get prior to their resettlement, the expenses which come with these facilities are not bearable for them. The land which is given to them for settling down, is not enough to earn to meet all those expenditures. The indigenous communities in Bakun went through challenges after being displaced and still working to sustain a quality life in their new place. The field surveys stated that being relocated due to dam those people were left with not much land to cultivate, and not so much jobs to earn their livelihoods. Along with the time, different support from local government helped them to make the new place their home.

Fearnside (2016) conducted a research work on the effectiveness of dam projects on the local people in Brazil. As dam construction involves so much of corruption, the monetary support from government to the displaced people never reaches. 2300 people were displaced due to dam building project in Amazonia who are now struggling to meet their basic needs. The decision of building dam to make life easier for protecting environment and saving people from the shortage of electricity gives no help to the people who get displaced for this purpose. The irony is that the voices of the people who suffer don't get heard by the officials so often.

Habich (2017) researched on how the negative impacts of constructing a dam can be reduced by government efforts. After taking into account the cost of displaced people, China government in May 2011 released a statement on raising the standard of living of these people. The government decided to increase local GDP by supporting the reallocated people by providing them enough chances to shift their economic condition by promoting community service activities so that the local per-capita income benefits come into action (Alam, Sultan, and Afrin, 2010). The government also helped local resource management and entrepreneurial activities through government management to which the displaced local communities could utilize their talents for income generating activities (Alam and Hossain, 2018). Yet increasing production of dams resulted in poor water resources for irrigation and local environmental concern (Alam, 2013) in nearby areas of the dam project. Even though government provides some beneficial support, the main occupation, cultivation, get affected adversely. Thus, the calculated costs overpowers the benefits of dam.

In 1979, a survey found that 69% of the Chakma felt that the Kaptai dam created the food and economic crisis for them. 89% of the Chakma people also said that they were displaced by the inundation of their homes and land. Almost 87% of them said that they had to face serious trouble in building new homes and 69% complained about insufficient compensation and corruption of government officials. 78% of them complained of having no opportunity for jobs in the hydroelectric project and 93% of them claimed a better life and economic condition before the Kaptai dam was constructed (Ibid., p. 29.)

Several studies have conducted research on the cost method of changes in land use. Empirical literature provides significant evidence on the cost incurred, specifically in terms of land use, due to construction of the dam. This existing literature would assist us in comprehending and analyzing our model results based on the methodology of our study.

One of the several studies that discusses about the cost of changes in land use due to the construction of reservoir by Tirkaso and Gren (2018). In the study, the authors estimated the cost frontier function of various projects on Hydropower-Related Biodiversity Restoration in Sweden, predicted the corresponding level of efficiency using a stochastic frontier analysis. The study uses dataset from 245 such projects in Sweden that have been conducted in the duration from 1986 to 2015. Corresponding to the biodiversity restoration projects, the twice-differentiable Cobb-Douglas frontier cost function holds the dependent variable to be the cost of biodiversity restoration, and the independent variables to be the ecological output, wage rate for labor, and interest rate. The results indicate an evidence of cost inefficiency in the projects, showing an average efficiency score of 55%, which suggests potential to minimize cost efficiency loss by 45%.However, if we want to look particularly for the cost of changes in land use, we can observe from the results of the study that while construction of natural fishways has been the most expensive measure, improving spawning grounds has been the least costly measure (representing around 17% of the cost of a natural fishway). This indicates that for the various projects on Hydropower-Related Biodiversity Restoration in Sweden, the cost of changes in land use has been the least costly measure amongst all other monetary and non-monetary costs, according to the results of this study.

Morimoto and Munasinghe (2005) studied on small hydropower projects and sustainable energy development in Sri Lanka. They discuss about the role of small hydroelectric power projects in sustainable energy development in Sri Lanka using a Sustainomics framework. This study uses multi-criteria analysis (MCA) to examine the impacts of hydroelectric projects, dealing with multiple objectives simultaneously, from a sustainable development perspective. They considered three main sustainable development issues that comprise the economic costs of power generation, ecological costs of biodiversity loss, and social costs of resettlement. This multi-dimensional analysis, which includes environmental and social variables, supplements the more conventional CBA that is based on economic values alone. In order to analyze the cost of changes in land use due to the construction of reservoir, the authors have used an environmental indicator, and termed it as biodiversity index. It reflects several key characteristics. One of its elements concerns the relative biodiversity valuation of the project, because the value of the area lost is a function of the proportion of the habitat that is lost. The index observes the correlation between the land area inundated and affected and the energy storage capacity. The results of the study found that there is little correlation between the quantity of biodiversity index, average generation cost, number of resettled people, and electricity generated.

Moreover, Hreinsson and Elíasson (2002) discuss on Optimal Design and Cost/Benefit Analysis of Hydroelectric Power Systems. To do so, they use a sequence of hydro-projects simultaneously rather than a single project, and perform the global optimization simultaneously on all projects using both a proposed one-shot approach as well as iterative techniques. They also use Lagrange multipliers as economic signals between stages in the expansion process, while at each stage in the parameter design and optimization of each hydroelectric project, the method of Genetic Algorithms (GA) is used. The study uses the project construction cost as a function of the parameters, such as reservoir size and dimensions, dam height, diameter of tunnels, penstocks, etc., and this total cost, consisting of construction and operations cost, is represented by the total discounted project cost NPV (net present value) using an appropriate discount or interest rate. The results found that in the project planning report, the size of the power plant and the size of reservoir is selected on basis of a power market scenario at the expected construction time of the plant, however the optimization assumes infinite demand. Nevertheless, the solution is not far from their feasible arrangement.

However, Gren and Amuakwa-Mensahdiscusse (2018) study the cost of land use due to the construction of reservoir while estimating the economic value of site quality for uncertain ecosystem service provision in Swedish forests. The aim of this study was to measure the impact of site quality on the mean and variability in the economic value of timber and carbon sequestration in Sweden. The study used a method for calculating the impact of site-specific ecological conditions in Swedish forests on the economic value of uncertain ecosystem services in terms of timber and carbon sequestration. An econometric analysis is conducted using panel data for a period over a 50-years on forest growth, management practices and site quality indicators for various forest regions in Sweden. Indicators for management practices include harvest of trees, fertilization, thinning and scarification, and a site quality index is used to measure the given environmental conditions at each of the forest site. The results demonstrate the importance of the forest site quality by showing that a marginal increase in site quality can raise the economic value of timber and carbon sequestration by 9%, whereas neglecting uncertainty can underestimate the value of the contribution by 12%. These results indicate that management practices that improve site quality have the potential of raising the total economic value of forest ecosystem and stabilizing its volatility.

From the above reviews of the specified impacts of the issues related to the Kaptai Dam, it is observed that there were no in depth study on the environmental cost benefit of this dam. Thus, this study aims to contribute into the existing literature with its findings into the academia in context of environmental cost benefit analysis of Kaptai Hydro-electric power plant.

## 4 Method and Methodology
### 4.1 Data Collection
The main sources of data are both primary and secondary. The secondary data has been collected from Bangladesh Porjoton Corporation, Statistical Year Book of Bangladesh, Kaptai Dam Authority, Fisheries Department and Ministry of Hill Tracts. However, collecting data from Kaptai Dam authority is very complex and thus electricity production data has not been collected. The primary data was collected through face to face interview for tourism and all other indirect costs including environmental cost, cost of displacement, cost of lives lost and cost of land use changes. Two separate structured questionnaires have been used for tourism, and indirect costs associated with the ethnic households. In addition to these, Key Informant Interviews (KIIs) were conducted to assess the average cost and price of fisheries caught. Table 1 below shows the sources of data including their nature.

**Table 0:1: Nature and Sources of data**

| Activity/Sector | Nature of data | Sources of data | Sample Size / Year included |
|---|---|---|---|
| Electricity production | Secondary | Statistical year book | 47 years |
| Production of fisheries | Secondary | Fisheries Statistics | |
| Tourism | Primary and secondary | Sample survey | 200 |
| Environmental costs including loss of households arable land, loss of lives and losses of resources | Primary and secondary | Sample survey | 202 |
| KIIs (Fisheries) | Primary | Face to face interview | 3 |

The tourism data has been collected through convenient sampling from the tourist spots in Rangamati. A well prepared questionnaire was used to collect the primary data. The primary data was collected through face to face interview to the tourists. The number of sample size of tourist data is 200. For the costs, another questionnaire was used to collect data from the households. The data was collected from both sides of the lake; Kaptai and Rangamati sadar areas through the face to face interview. In addition, secondary data was collected from different NGO's, Govt. officials and local community leaders to finalize the report.

## 4.2 The Model

Since Kaptai dam is providing 5% electricity demand of the country, we can find the total direct benefit by valuing the electricity supply of the dam in monetary term. Both producer surplus and consumer surplus will be measured through the direct benefits. On the other hand, indirect benefits are fish culture, tourism and recreation which will be calculated in third party gains. In estimating the cost, we will consider the direct cost of producing electricity which includes initial fixed cost and variable costs in the following years, but the indirect costs like loss of agricultural/cultivable land, loss of houses/accommodations of the local people, security cost of the government and the value of lives lost in the area because of instability (locally named as Guerrilla War). For calculating the environmental costs and benefits of the Kaptai Dam we consider the value of electricity, the value of tourism and the value of fisheries in benefit side. In contrast, the cost of installation of the dam, the costs of electricity production, the costs of environmental degradation, the cost of security, cost of displacement, value of lives lost are considered as the cost of the dam.

Thus the net benefit of the dam stands as follows,

$$NB = (Net\,Value\,of\,electricity + Net\,Value\,of\,fisheries + Net\,Benefits\,of\,tourism) - (Cost\,of\,security + cost\,of\,displacement + Cost\,of\,land\,use\,change + \cos t\,of\,lives\,lost) ------(1)$$

The issues used in equation 1 have been estimated using the following methods;

### 4.2.1 Value of Electricity

The value of electricity is the product of electricity and its unit price. In terms of economic theory it is revenue (R) which is the product of total number of output (Q) and its unit price (P) which is a direct method. The electricity also includes its production costs. Therefore the net present value of the electricity is the additional gain of the electricity production. The formula for net present value stands as,

$$\sum_{t=1962}^{2020} NB_t = \sum_{t=1962}^{2020} Q_t *(P_t - C_t) ---------------(2)$$

The total capacity of the dam is 230MW though the average production is about 181.25 MW because of the low production during the dry season (April-June). Therefore, to generalize the production of the Kaptai dam we will consider 180 MW/ day/average. Since the total production is the daily data we will convert it into the yearly data by multiplying by 365 day. There is one problem to estimate the total net value of the electricity production due to the unviability of the long run data of electricity production and its price over the periods. Therefore, the maximum capacity of the production has been used with its mean value. Similarly, to address the problem of unavailability of the price data Consumer Price Index (CPI) has been used to convert the yearly data. To calculate the gain from electricity we consider the present unit price (per 1 kwh) in BDT 7.78 (CosttoTravel, 2021) and unit cost BDT 4.20. The price of the electricity is available in Bangladesh but the cost data is not. Therefore global average of hydroelectricity production cost $0.05 (The energy factbook, 2021) has been used as a proxy. The dollar value of the cost has been converted to BDT with

exchange rate BDT 84. Later the net benefit of 2020 has been discounted using the consumer price index of Bangladesh. Using 2010 as the base year (Knomea, 2021) consumer price index for 1986-2020. But, to calculate the net benefit of the electricity till 1962, therefore the rest of the years CPI were simulated following previous years.

4.2.2 Value of Fisheries

Valuing the fishing of Kaptai lake, also depends on the direct method which is price times the quantity. Kaptai Lake is contributing directly to national fish production. Recent statistics shows that the lake has produced about 9,500 tons of fishes during 2016-17 fiscal year amounting Tk 120 million (Financial Express, February 07, 2018). That is average price of per kg fishes of the lake was TK 126.32. Surprisingly, the KI interviews with the fishermen explored that the average price of per kg of fish ranges between 66.67 to 75 taka which is very low compared to the market prices. The major reason of this low price is because of the advance loan taken by the fishermen that is locally known as 'Dadon'. Once the fishermen borrow 'Dadon' from the whole seller/Aratder, they are supposed to sell the fishes they catch to the respective seller/Aratder. The fishermen also admitted that the price they sell to whole seller/Aratder is low compared to market price. Therefore, the average price of the fishes will be used as of taka 126.23 to value the fisheries of the remaining years. To convert the price of the fishes, I will use CPI of the corresponding year and multiply the yearly production with the value of the CPI. So far we have total fish production data from 1983-84 fiscal year to 2017-2018 fiscal year, and revenue data of the lake for 2006-2007 to 2017-2018. Now, using this information, we have converted the previous years' values and sum to estimate the total value of fisheries. For conversion of the prices, we use yearly CPI.

The value of fishing is estimated through the 'use value' method. Fishes are caught and sold in the market. For valuation of the fisheries valuation techniques that are mostly used are; (1) conventional economic valuation that uses households survey for economic cost-benefit analysis; (2) economic impact assessment to estimate output volume and value on the basis of price through monitoring fish markets; and (3) socioeconomic and livelihood analysis (World fish, 2008). Among these three methods, the impact assessment is found to be useful and effective as it requires little data (World Fish, 2008). Likewise the world fish valuation methods (2008), we will use the impact assessment. Here, we will use cost benefit analysis to value the contribution of the fisheries caught from Kaptai Lake. The model that we use is as follows;

$$NPV_{fish} = \sum_{t=1986}^{2019}(R_t - C_t)(1+r)^t \quad\text{-----------------(5)}$$

Here, Bt refers the total benefit or revenue of the fisheries that is calculated by multiplying the yearly catch by the average price of the fisheries. Here r refers for discount rate. Similarly, Ct refers the cost of the catch of the fisheries. Here, we have fishery catch data but cost of the catch is not available as yearly data and thus for this data we use a proxy of this by a key informant interview (KIIs) with the fishermen. The collected data through KIIs, we have extrapolated for the yearly data using a discount factor. The Key Informants (KIIs) opined that per KG fish costs 15 taka whereas the sell value of per KG fish is 75 taka with Dadon.

Here, R is total revenue earned by each year, C is the cost of catch, is a discount rate which would be 10% by assumption, and t refers for year. The secondary data of fish production that has been collected from Department of Fisheries, Rangamati was for 12 years (2006-07 to 2017-18) only. Thus for missing year data, we will simulate the present data for previous years (1986 to 2018) with the help of CPI (Consumer Price Index). Regarding the cost of the fisheries, the labor cost, food cost and transport (boat used for fishing) were considered. The total cost was thus calculated based on total yearly production, as a whole. The total cost is subtracted from the total value of the fishes caught by every year to assess its yearly net value.

**Table 0:2: Yearly Production and Revenue of fisheries**

| Fiscal Year | Production (Ton) | Revenue in M BDT | Price/KGg (BDT) |
|---|---|---|---|
| 2006-07 | 5389 | 288.87 | 53.60364 |
| 2007-08 | 7633 | 423.63 | 55.4998 |
| 2008-09 | 5495 | 314.49 | 57.23203 |
| 2009-10 | 7115 | 494.99 | 69.56992 |
| 2010-11 | 8974 | 626.96 | 69.86405 |
| 2011-12 | 8421.75 | 694.33 | 82.44486 |
| 2012-13 | 8813.56 | 766.58 | 86.97734 |
| 2013-14 | 7725.55 | 668.99 | 86.59448 |
| 2014-15 | 8644.85 | 867.75 | 100.3777 |
| 2015-16 | 9589.6 | 996.34 | 103.898 |
| 2016-17 | 9974.44 | 1203.32 | 120.6404 |
| 2017-18 | 10140.78 | 1242.5 | 122.5251 |

Source: Rangamati Fishery Office

### 4.2.3 Valuing of Tourism of Kaptai Lake

To value the tourism in Kaptai Lake, the zonal travel cost method to estimate the visitation rate. The travel cost method in general is used to estimate the benefits derived from environmental recreation sites such as forests, fishing and hunting spots etc. while the zonal travel cost (ZTC) model is used to find out the link between travel costs (tc) and visitation rate (V). The econometric model for this can be written as:

$$Vi = \alpha + \beta_1 TC + \beta_2 MI + \beta_3 Alone + \beta_4 Dhaka + Ui -----(6)$$

In the above equations,

Vi refers visitation rate; this is the dependent variable, TC refers travel costs, MI refers monthly income of the respondents, Alone (Dummy variable) refers the individual is alone (=1), or in a group (=0), Gender (Dummy variable) refers male if Gender =1 and female if gender =0, Dhaka is the zone variable and the value of Dhaka =1 refers that the respondent is from Dhaka and 0 for otherwise, and Ui refers for error term.

The visitation rate (Vi) is calculated through dividing the total number of visit per zone by total potential number of visitors in the zone. The potential number of visitors are the population of the zone whose income exceed the mean income of the visitors. So, the visitation rate refers per 1000, or per 10,000 or per 100,000 from each zone. In the current study, the visitation rate was calculated for 1 million people. That is per million how many people have visited Kaptai Lake in one year.

### 4.2.4 Environmental Cost

To estimate the cost of environment, the Contingent Valuation Method (CVM) is used. The environmental cost includes losses of environmental services and losses of households' displacement. It is estimated that about 100,000 people have been displaced from the dam area due to its construction and production. The CVM for Environmental Cost has been used which is as follows;

$$EC = \sum Monetary\ values\ of\ losses\ of\ environmental\ services ----------- (7)$$

### 4.2.5 Security Cost

After the dam was built the hilly people started fighting against the government. Therefore many security forces were employed in that region. Therefore, direct cost of security forces were considered as the cost of security in the Kaptai dam are. For the security cost yearly costs of the security is considered as additional security costs of the dam.

### 4.2.6 Value of Statistical Lives

Every life has a value. Usually, the value of statistical lives is calculated using the following formula;

Value of life= (Reduction is wage/reduction in probability of death)

While calculating the value of life, it is assumed that all jobs; high risky jobs and low risky jobs, are identical. Since the data collection of this study did not consider the job types, therefore a simplest form of calculation has been used to calculate the value of life. In this study the value of life refers the opportunity cost of a deceased life in terms of monetary income. To make it simple, this study considers that the working life of a person is T years and his/her annual earnings is W taka. If the person dies at year X, then the opportunity cost of dying the person is (T-X)*W which is simply the money income lost for his/her family.

## 5. Discussion and Results

### 5.1 Descriptive Analysis

For valuing Kaptai Lake and its services, it is important to know the socio-economic condition of its users and service receivers. Therefore, socio-economic background for tourists is important for valuation of the tourism service and socioeconomic background of the surrounding people is important to estimate indirect costs associated with the dam construction.

### 5.1.1 Description of Tourists

For valuation tourism service of Rangamati, we have studied the socio-economic background of the visitors. Based on the data from the survey, the following pie-charts have been constructed. We can get a general idea about the occupation, gender, age, marital status and area of residence of the visitors from these charts:

First of all, considering the area of residence of the visitors we can see that the majority of the visitors were coming from Chittagong, Dhaka, Rangpur, Rajshahi and Sylhet divisions in chronological order. As majority of the visitors were from Chittagong and as Rangamati is included in the Chittagong division itself, it can be understood that most of the visitors were local residents.

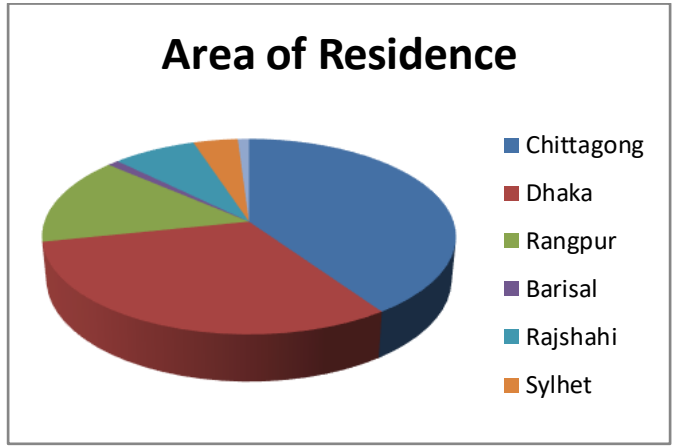

**Figure 2: Locational Distribution of the Tourists**

Secondly, in terms of educational qualification, it was seen that most of the visitors finished until either secondary or graduate levels. This means, that most of the visitors were educated beyond the primary level.

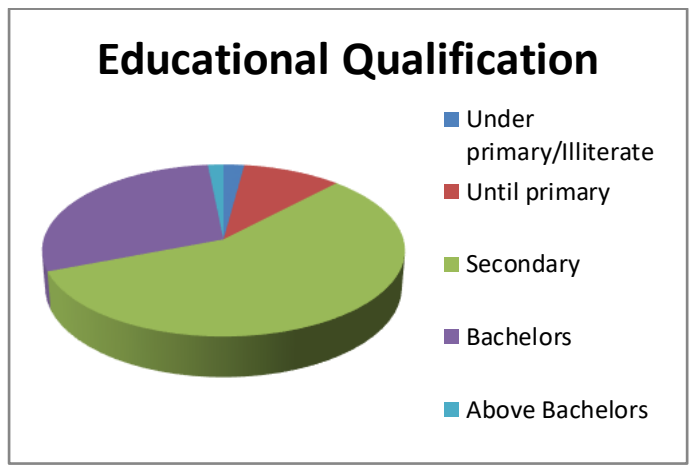

**Figure 3: Educational Qualification of the Tourists**

Thirdly, when it comes to occupation, we can see from the following pie-chart that most of the visitors were involved in business, service or other professions. Very few people were doctors, engineers or dependent. It can be assumed from this data that the respondents of this sample surveyed belong to the middle class group of the society.

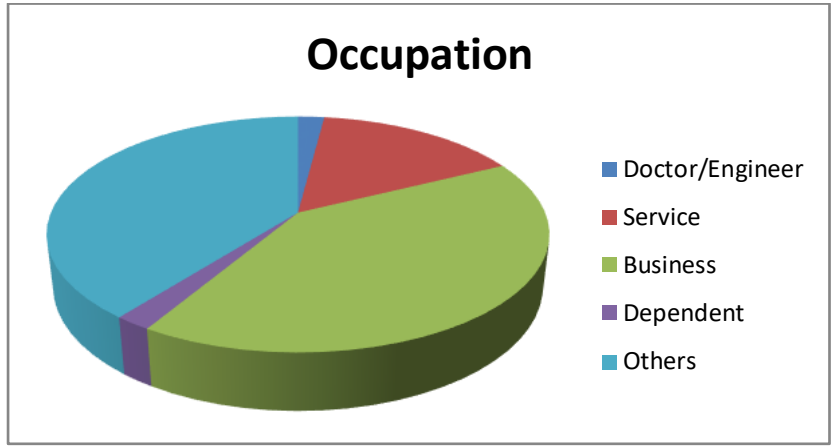

**Figure 4: Occupation of the Tourists**

Fourthly, when it comes to age, it can be seen from the following pie-chart that the visitors were very young as majority of the visitors were aged less than 25. This statistics is very meaningful as most of the visitors in Rangamati are students in groups.

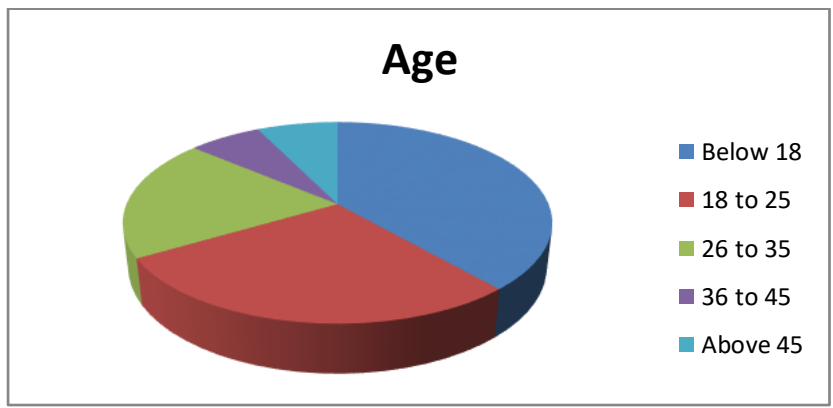

**Figure 5: Distribution of Age of the Tourists**

Fifthly, when it comes to gender, it can be seen that the visitors were mostly male. This gives us an insight into the gender-based societal restrictions prevalent in Bangladesh as women are not given much liberation to travel on their own.

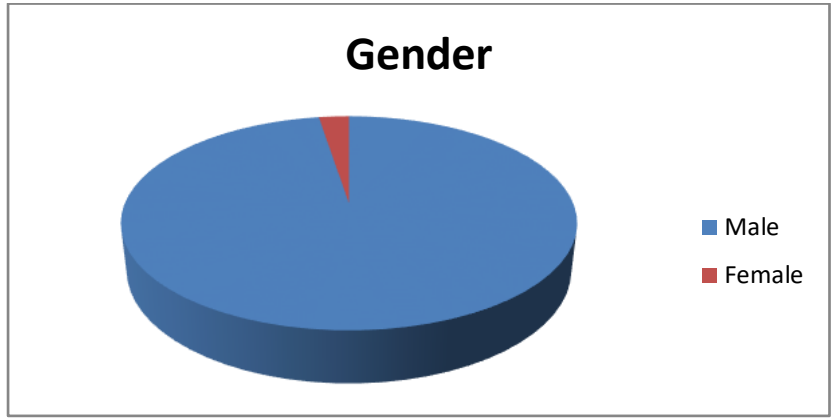

**Figure 6: Gender Distribution of the Tourists**

Sixthly, in terms of marital status, it can be seen that most of the visitors were either married or unmarried. None of them were divorced or widowed.

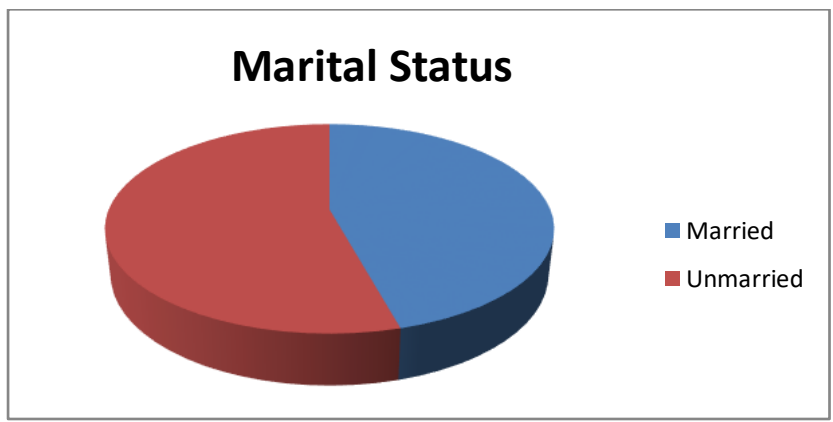

**Figure 7: Marital Status of the Tourists**

Now, to understand the economic standing of the visitors, we can look at the following table of summary statistics of their monthly income and total family income:

**Table 3: Summary Statistics of Income**

| Variable | Mean | Standard Deviation | Minimum | Maximum |
|---|---|---|---|---|
| Monthly income (Taka) | 20151.49 | 20858.8 | 1800 | 200000 |
| Total family income (Taka) | 31470.87 | 29145.66 | 2500 | 200000 |

It can be observed from the above chart that most of the visitors belong to the middle to upper-middle income class of people in the society.

Now, let us look at whether the visitors came alone or not. It will give us an idea about their visitation preference and their travel cost to Rangamati. The respective series is given below:

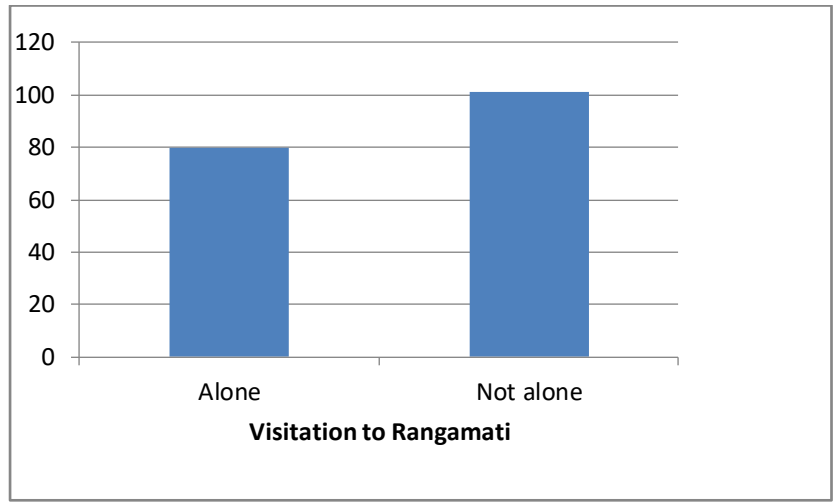

Figure 8: Preference of Visitation

It can be seen that most of the people were not alone during their visit to Rangamati. This means that the visitors have more possibility of sharing their travel cost with their other people throughout the course of the visit.

Now, it is important to understand whether the visitors were satisfied with their visitation to Rangamati. The following diagram can make it clear:

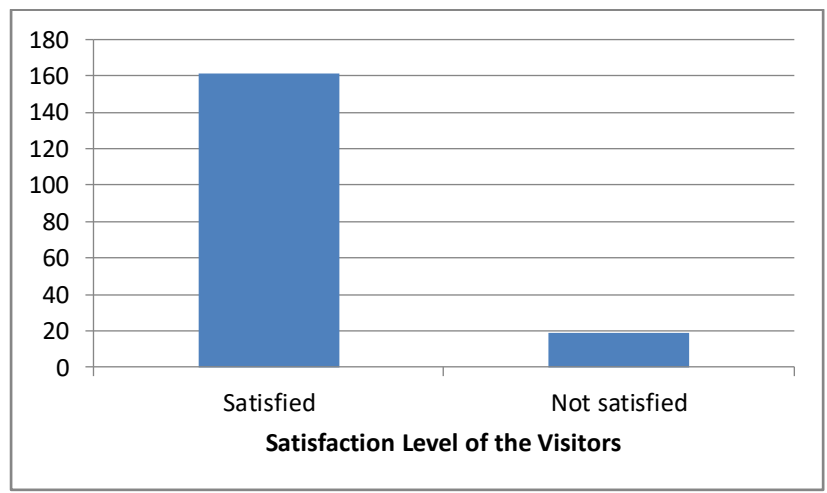

Figure 9: Satisfaction Level of the Visitors

So, it can be seen that most of the visitors were satisfied with their stay at Rangamati. The next important thing to be identified here is the criteria which made the visitors satisfied with their visit to Rangamati. The following series can make it understandable. Such as:

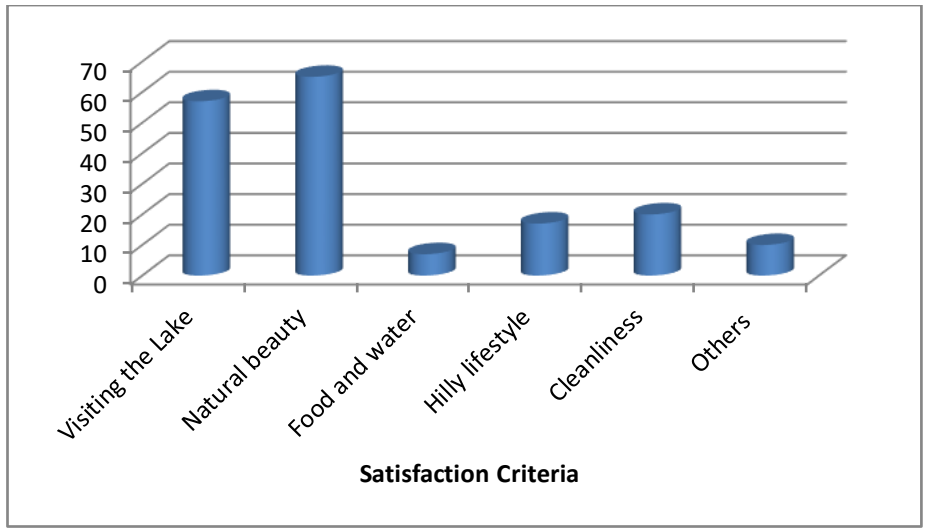

**Figure 10: Satisfaction Criterion of the Visitors**

It can be seen that the visitors were most satisfied for visiting the lake and exploring the natural beauty of the hills and the lake.

This brings us to the most important part of the research. We need to know the opinion of the visitors regarding the services provided at Rangamati and whether they want to improve it or not. We can understand the perception of the visitors through the following pie-chart and bar diagram:

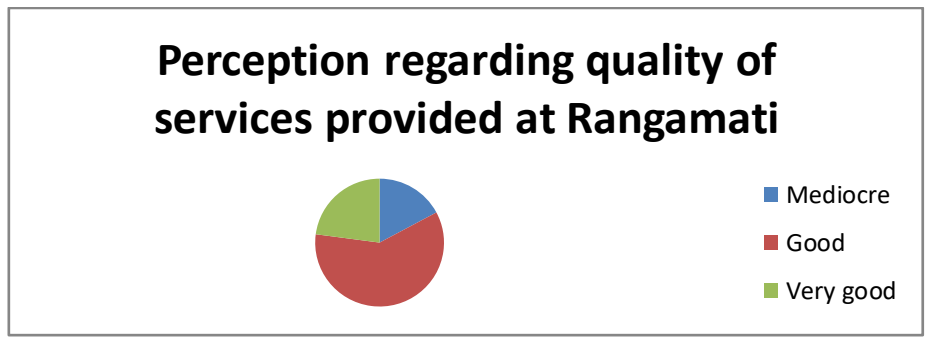

**Figure 11: Perception about the Quality of Services**

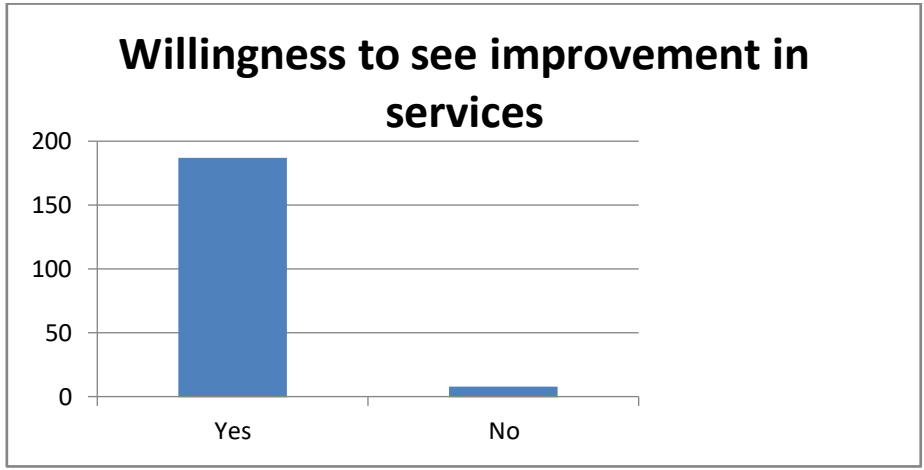

**Figure 12: Respondents Willingness to see the improvement of the site**

We can see a very contradictory picture from the above representations. That is, even though majority of the visitors rate the quality of services provided at Rangamati as good or very good, they are still willing to see more improvement in the current situation.

Now, the question is whether the visitors want to take part in improving the services by contributing/investing in the government projects. Surprisingly, most visitors were unwilling to invest for the cause. Such as:

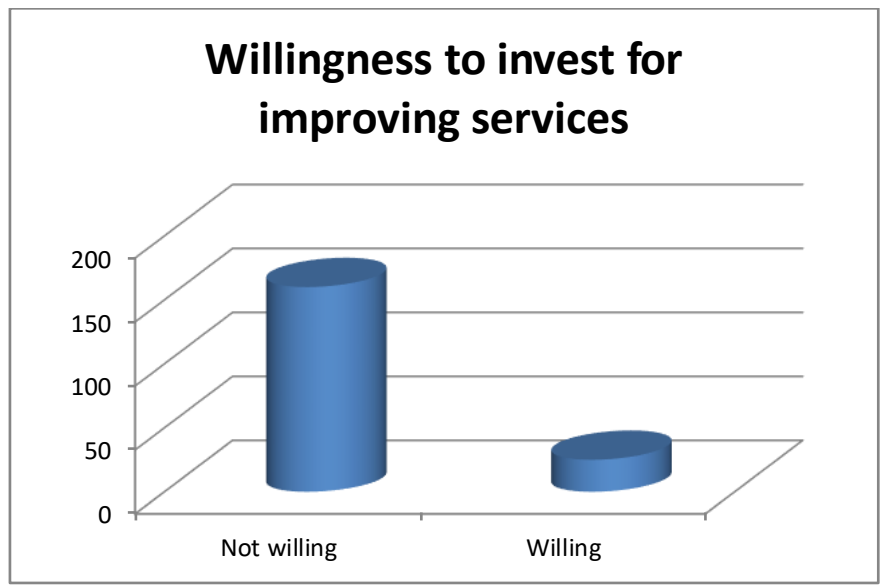

**Figure 13: Respondents Willingness to Invest for Services**

Again, when asked whether the visitors would like to visit again if better facilities are provided for sanitation, security, environmental conservation, residence and transportation purposes, all the visitors responded positively/affirmatively. All the visitors also expressed that they would like to visit more often if the entrance fees imposed by the government reduces and their own personal income increases.

From the above analysis of the sample data, we can understand a lot about the socio-economic status of the visitors, their visitation perceptions and preferences in the future towards Rangamati.

### 5.1.2 Descriptive statistics of fishermen

To accomplish the value of fisheries, 3 Key Informant Interviews (KIIs) were conducted with 2 fishermen who are also owners of the boat; the main fishing effort and 1 fisherman work daily basis. The KIIs findings show that the average age of them is about 32 years and their average family size is 5. The fishing in Kaptai Lake allows 9 month fishing to them. On average the fishermen catch fishes in the whole allowable season (9 months). In each of the boat on average 6 fishermen catch fishes and their average daily fishing is about 360 kg. The two owners mentioned that they got advance loan from the whole seller/ Aratder and therefore they are bound to sell their catches to them. They also mentioned that the selling price to the whole seller/aratder is comparatively low. The average selling price is 75 taka per kg and average daily cost of the fishing per boat is 5433 taka. However, the average selling value of the `per boat fishes is about 28500 taka.

### 5.1.3 Description of Community People (Related to Costs)
#### *5.1.3.1 Age, Gender, and Educational Background of the Primary Respondents*

A good number (202) of respondents have been reached for the purpose of interview and as a source of primary data. The data reveals that the average age of the respondents is 51 year, and majority lies in between 51 and 64. It implies that majority respondents are of course adult and parents. However, the lowest age is 24 (far above 18) and the highest age is 90. In case of gender, around 95% respondents are male. It is quite a gender bias data. However, it is justified in the ground that the issue of the research is an external (not household) one, and also male peoples are require more to involve in conflicting issues. It is a similar trend with our national picture, where it is a patriarchy society and a males lead families . If we look into the educational background, it is quite improvising. All the respondents are literate, whereas national literacy abruptly is 23%. The average education level is class VIII (secondary school) with the highest graduate with 16 years of schooling and the lowest is class II, having alphabetization skills. As the topic is a multi-dimensional livelihood issue majority respondents have directly experienced through their life. Hence, they have good familiarity over the dimensions and nature of the topic. Among total respondents almost 96% are ethnic minority of the region and remaining 4% are mainstream Bengali people.

*5.1.3.2 Family, Income, and Household Issues*

We could reach the family data of 96% respondents. Among them average family has 4 children, implies that average family size is 6. It is a bit higher than that of the national scenario where majority family has 2 children. As a government of the most populous country Bangladesh government also promotes that having one child is good and having 2 is enough. Moreover, the more the family is educated and living in a urban areas, the more they have concerned on the children and their nutritional outcomes (Islam, Alam, & Afzal, 2021). However, the externality of the data shows that the highest number of members in a family is 10. It can be the issue of joint family culture still exist in the ethnic minority community of the Chittagong Hill Tracts Region of the country. Data also reveals that on an average, 3 children goes to educational institutes. The rate is quite high. On the other hand, explanation of the absentees is that either they are underage to start school neither dropout mostly for very remote territory. Interesting is that though they somehow suffered (will explain details later) for their livelihoods still their average family income (16,895 BDT) is much higher than that of the national average (11,000 BDT) calculated in 2011 (ref.). In case of consumption, average is 14,645 BDT, what is quite higher than that of the national (ref.). One of the significant issues of consideration here is that, the area is very remote and mode of transportation is very tough. In most cases they have to use 3 hours river journey through water vessel to come to district head-quarter town Rangamati. The dependency is acute for the medical treatment, and higher education purpose. In question of electricity consumption, it has found that 72% of the household has electricity connection and their average monthly bill is around 618 BDT. However, still 50% of them have monthly savings of around 3,650 BDT, what is still good figure for this society.

*5.1.3.3 Dam, Hydro-electric Project, Consequences, and Rehabilitation Issues*

There is high awareness on the hydro-electric project among the respondents. More than 99% respondents have positively replied that they have clear understanding on the issues of dam, hydro-electric generation, consequences on the livelihoods, government's rehabilitation efforts, and surrounding environments. On several particular issues of land there are substantive opinions. Around 99% respondents have replied that they have lost one or other kind of land. Such as, respondents have lost on an average 207 decimal of household/homestead, 720 decimal of agriculture, 235 decimal of forest, and 1018 decimal of total land respectively. As an immovable property land has significant value in the non-industrial society. Hence, the loss of land is highly significant in their lives. However, in case of inquiries on the loss of cattle and pond land, only a few suffered. The research has considered the as insignificant.

In case of getting supports from the rehabilitation efforts of the government only around 28% respondents have received in the form of land and/or cash money. Only 18% respondents have got as an replacement on an average 357 decimal of land and/or 16% have received on an average 1,941 BDT as cash money. However, the highest the value of land rehabilitation is 2000 decimal, where the highest value of cash support received is 16,000 BDT. Strikingly limited (6%) respondents are satisfied with the rehabilitation arrangements, on the other hand 94% are thoroughly dissatisfied and experienced serious discomforts in their livelihoods.

*5.1.3.4 Valuing the Hydro-electric Dam/Cost-Benefit Analysis of the Hydro-electric Dam*

The estimation of the exact value of the cost and benefits of the hydro-electric dam cannot be possible with 100% accuracy. The challenges of 100% accuracy includes valuing human lives, excluding/including natural disasters and many more. Though this research has come up with a quantitative and mathematical analysis, but it can be treated as great source of understanding. Issues of valuation includes production and price of rice, crops, fruits, fishes, fire-wood, and others. Those can be seen at a glance in the following table.

**Table 4: Descriptive Statistics of the Agro-based Production in the Dam Area**

| Sl. | Name of Item | Average Annual Production | Season | Average selling value (BDT) | Total price | |
|---|---|---|---|---|---|---|
| 1 | Rice | 326 Mound | 2 | 1050* | 342,300 | |
| 2 | Crops | 1375 KG | 1 | 761 | 10,046,375 | |
| 3 | Fruits | 745 KG | 1 | 636 | 4,73,820 | |
| 4 | Fishes | 654 KG | 1 | 350 | 2,28,900 | |
| 5 | Wood | 180 Mound | N/A | 4,82,238 | 8,68,02,840 | |
| 6 | Medicinal Plant | 18 | 1 | 100 | 1,800 | |
| 7 | Acquired Land | 1018 Decimal | N/A | 18,036 | 1,83,60,648 | |

*Source*: Calculated from the survey of this study, *Dhaka Tribune, 19 May, 2021

*Note: The price of rice that the respondents reported was excessive and thus the price of rice (1050/mound) considered for the calculation was a secondary information*

### 5.1.3.5 Human Causalities and Some Other Issues

On an average 2% of the respondents and their family members have got injury/death due to the conflict arose from the dam issue, who belongs to the average age of 35 years. Also, 1% experienced various forms of mental tortures. Valuing human causalities and injuries are challenging as because every human being could have unexplored potentialities in an earthly life. In the developed countries a common practice to value these issues is the value of statistical lives. However, in Bangladesh no evidence was observed which used the value of statistical lives. Therefore, opportunity costs of early death has been estimated considered the income as a proxy variable.

### 5.2 Econometric Model

### 5.2.1 Visit generating function

After regression, the results obtained for the visit generating function is summarized below in Table 5.

**Table 5: Regression Results for Tourism**

| Variables | Model | | | |
|---|---|---|---|---|
| Dependent variable | Visitation rate | | | |
| Explanatory Variables | Co-efficient | S.E | T | P > \|t\| |
| Travel cost | -0.0210306 | 0.064171 | -3.28 | 0.001*** |
| Monthly income | .0014136 | .0024446 | 0.58 | 0.564 |
| If alone | 31.34816 | 76.24528 | 0.41 | 0.682 |
| Dhaka | -387.2958 | 58.05345 | -6.67 | 0.000*** |
| $R^2$ | 0.2542 | | | |
| F (4,116) | 21.16 | | | |
| Prob > F | 0.001 | | | |

Here, ***$p \leq 0.01$.

Table 5 shows that if the travel costs increase for one unit visitation rate decreases for 0.02 units which reflects the 2% changes. The R-squared explains that 25% variation in the visit generating function is explained by the regressors and the model is significant which is verified by the F-statistics which is significant at below to 1% level of significance.

From the regression results above, it is seen that the monthly income and travelling alone are not significant in explaining the variation in visitation rate. But the travel cost and Dhaka dummy are significant in explaining the variation in visitation rate. Here, the travel cost and Dhaka dummy are significant at 1% level.

The a-priori expectation was that the travel cost and travelling alone dummy will show negative co-efficient signs whereas the monthly income and Dhaka dummy variables will show positive co-efficient signs.

But, only $β_1$ and $β_2$ are found with expected signs. These mean if the travel cost increases by 1 unit, the visitation rate decrease by 0.0210306 unit. If monthly income increases by 1 unit, the visitation rate increases by 0.014136 unit. If the visitor travels alone, then the visitation rate will increase by 31 unit more if visitor visits with a group. If the visitor is from Dhaka, then the visitation rate will decrease by 387.2958 units

### 5.2.1.1 Economic Valuation of tourism:

After conducting the regression, we finally find out the economic valuation of Rangamati zone-wise. We then find out the maximum willingness of the visitors to pay for site visitation and thus find the net economic contribution of the visitors to Rangamati. It was found that the visitation to Rangamati contributes 289.71 Million BDT (approx.) or 3.62 Million USD (approx.) per annum/year in Bangladesh. The following graph depicts findings:

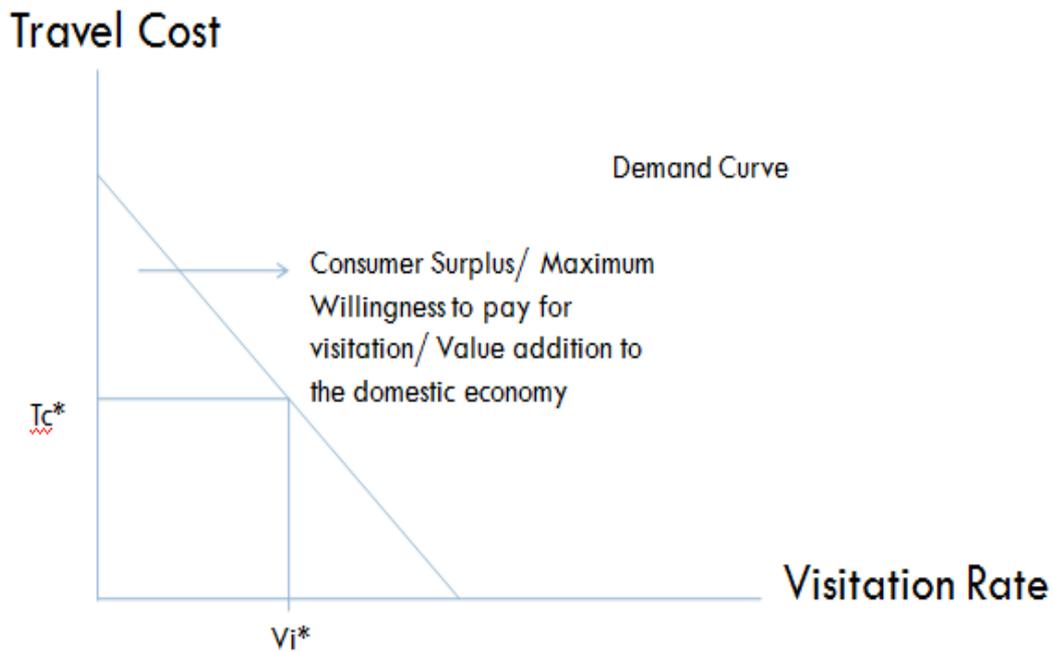

**Figure 14: Theoretical Visit Demand Curv**

The annual value then has been converted for all the sample years using the Consumer Price Index (CPI) using base year 2010. After converting the value of tourism for all sample years, the cumulative sum of the tourism stands for BDT 20070 M.

5.2.2 Value of Fishing

The value of fishing is estimated using the use value method. Fishes are caught and sold in the market. For the fisheries, valuation techniques that are mostly used are; (1) conventional economic valuation that uses households survey for economic cost-benefit analysis; (2) economic impact assessment to estimate output volume and value on the basis of price through monitoring fish markets; and (3) socioeconomic and livelihood analysis (World fish, 2008). Among these three methods, the impact assessment is found to be useful and effective as it requires little data (World Fish, 2008). Likewise the world fish valuation methods (2008), this study used the impact assessment. Here, cost benefit analysis was used to value the contribution of the fisheries caught from Kaptai Lake. We have catch data from 1986-87 to 2019-19 (with extrapolation) and the market value and costs of catch by which we can estimate the total value of the fisheries. The model that we use is as follows;

$$NPV_{fish} = \sum_{t=1986}^{2019}(B_t - C_t)(1+r)^t \quad \text{------------------(5)}$$

Here, $B_t$ refers the total benefit or revenue of the fisheries that is calculated by multiplying the yearly catch by the average price of the fisheries. $r$ refers for discount rate. Similarly, $C_t$ refers the cost of the catch of the fisheries. Here, we have fishery catch data but cost of the catch is not available as yearly data and thus for this data we use a proxy of this by a focus group discussion and key informant interview (KIIs) with the fishermen. We have extrapolated for the yearly data using a discount factor. The discounted net benefit of the fisheries shows that since 1986-87 to 2018-2019 fiscal year the fisheries from Kaptai Lake has contributed BDT 33366.82 M.

5.1.2 Value of Electricity

After converting the net present value of electricity production for 2020 to the previous years, the cumulative sum of net benefits of the electricity for 1962 to 2020 stands for BDT 138341.7 M. To convert the previous year values to current year, 7% discount rate is considered.

5.2.4 Costs of Construction of the Dam

*5.2.4.1 Cost of Building and Maintaining the Dam*

The dam construction had establishment cost and compensation cost for land acquisition. The costs of building the dam is given in the following table:

**Table: 6 Cost of the Dam**

| Items | Amount (BDT/Rs) | Rate | Kind/explanation |
|---|---|---|---|
| Establishment costs | 2403M Rs | | Lump sum |
| Compensation costs | 37.8 M BDT | 700 BDT/Acre | Lump sum |
| Total Costs | 2440.3 M | | |

*Source: Calculation is made based on previosu*

The direct cost of dam construction was BDT/Rs 2440.3 M in 1957. Therefore, using the CPI (2010 as the base year) the present value of the direct cost of dam construction stands for BDT 404882.6 M.

*5.2.4.2 Cost of Land Use Change and Displacement*

The cost of land use change and displacement was estimated using the primary data collected through contingent valuation approach. The data shows that on average 1018 decimal land has been destroyed including forest land, arable land and homestead. According to the respondents the average value of land lost per respondents was BDT 17678 at then price. Therefore using the secondary information of land the lost is 708663 taka. In calculating the present value of the land lost and displacement was use of the following formula;

PV= Land Lost (1957 value)* (2019 CPI/1957 CPI).

As total number of displacement is 18000 families, the total cost of lost land and displacement stands BDT 12756 M.

*5.2.4.3 Cost of Controlling Insurgency and Maintaining Security*

The cost of security could be calculated by adding the total actual cost of security forces in the Hill Tracts. However, this data couldn't be collected due to the sensitivity of the nature of this data. This could be one major limitation of this study.

*5.2.4.4 Cost of Lives Lost*

According to the respondents of the survey, 3 of their family members died and one of the family member got injured in the conflicts between the militants and the government forces. The average years of the dead people is 35 Years. Among these 3 dead, two of them died in 1987 and another one died in 1994. According to Bangladesh Life Expectancy 1950-2021 (Macrotrends, 2021), life expectancy in 1987 was 56 years and in 1994 was 61 years (cited from United Nations- World Population Prospects). Therefore, one families lost 26 years and 16 years company and opportunity cost of their family members who died in 1987 and 26 years who died in 1994. With simple calculation respondents have average cost of live lost is 667690 for families who lost their member in 1987. However, the cost of lives of the families who lost their member in 1994 is 432270. Therefore, the average cost of lives lost is 366654 taka. There is no authentic data of total lives lost during the insurgency between the Shahti Bahini and the Bangladesh army. But, the government claims that from 1980 to 1991 Shanti Bahini killed about 1,180 people and kidnapped 582 people (Kabir, 2005). To calculate the value of lives lost this information of death has been used and thus the total loss of the value of lives stands for BDT 435.65M.

## 5.3 Estimation and Discussion of the Net-Benefits

The importance of the Net-benefits of the kaptai Dam is huge. Based on the findings of the separate benefits and indirect costs (external cost) of this study, the net-benefits (including direct and indirect costs) of the Kaptai Dam has been estimated as follows;

$$NPV_{Kaptai\,Dam} = NPV_{electricity} + NPV_{Tourism} + NPV_{Fisheries} - (PV_{Displacement} + PV_{Cost\,of\,lives}) ---------6$$

Now, inserting the net present values from above estimated benefits and present value of associated direct and indirect (external) costs, the net present value of the Kaptai Dam stands for,

$$NPV_{Kaptai\,Dam} = 138341.7M + 20070M + 33366.82M - (12756M + 432.65M)$$

$$= (191779.03 - 13188.65)M = 178590.38\ M$$

The estimated value shows that the net present value of the Kaptai Dam is 178590.38 M. In terms of profitability the dam seems a viable for Bangladesh. However, if all external effects are well calculated based on true data then it will show the real scenario. Therefore, calculating all external costs with real data may value the contribution of the dam.

## 6. Conclusion, Recommendations and Limitations

The net present value of the Kaptai Dam shows that it contributes tremendous value to Bangladesh. As a source of hydroelectricity, the Kaptai Dam is a source of clean energy, and its value might have been worthy of this dam produced a significant portion of the electricity. However, providing less than 5% of the national demand for electricity followed by various external and sensitive costs, the dam hardly contributes to the Bangladesh economy. Therefore, it is the right time for the country to think about the viability of this dam shortly if it successfully can produce more electricity by installing more solar panels, windmills, and nuclear energy. The dam might have been profitable if there were no ecological and political conflicts.

This study thus recommends that the country looks for other clean energy sources that have no chances of eco-political conflicts. For example, installing windmills in coastal areas and building solar panels in sunny fallow lands can produce more clean energy and create any external effects, unlike Kaptai Dam. However, this study suffers from various flaws like unavailability and access to public data. For example, the cost associated with security enforcement is not accessible, and yearly data for electricity production is not public (please see appendix). Therefore, all components of the expected model have not been estimated because of a lack of data. Despite these limitations, this study has tried its best to incorporate complete information to have a clear footprint of Kaptai Dam in terms of its net contribution to the country.

Therefore, considering all limitations, the present study could be a rough estimate of this dam which can help and be an inspiration to pursue a further study on the Kaptai Dam in the future.

World Fish, 2008. World fish Brief 1890: Tropical river fisheries valuation: Establishing economic value to guide policy, date accessed March 15, 2019. Available at http://pubs.iclarm.net/resource_centre/WF_1106.pdf

# 8. Appendix

These tables show the percentage of missing information, and actual and imputed values.

Table A1: Percentage of Missing points

| Variable | Actual | Data imputed | Generated by |
|---|---|---|---|
| Electricity | 0% | 100% | CPI |
| Tourism | 1.6% | 98.4% | CPI |
| Fisheries | 94.3% | 5.7% | CPI |
| Fisheries revenue | 31% | 69% | CPI |
| CPI | 60% | 40% | CPI |

TableA2: Actual and imputed values

| Variables | Actual Values | Imputed values |
|---|---|---|
| Electricity | - | 1962-2020 |
| Tourism | 2018 (survey), and 2017 (visitation data) | 1962-2016, 2019-20 |
| Fisheries | 1983-84 to 2017-2018 | 2018-2019 to 2019-2020 |
| Fisheries revenue | 2006-2007 to 2017-2018 | 1983-84 to 2005-06, and 2018-2019 |
| CPI | 1986-2020 | 1962-1985 |